\documentclass[manuscript,screen,nonacm,natbib=true]{acmart}

\usepackage{booktabs}
\usepackage{tikz}
\usetikzlibrary{positioning,arrows.meta}
\usepackage{listings}
\usepackage{xcolor}

\AtBeginDocument{%
  }

\setcopyright{none}
\acmConference{}{}{}
\acmYear{}
\acmDOI{}
\acmISBN{}
\renewcommand\footnotetextcopyrightpermission[1]{}

\lstdefinestyle{code}{
  basicstyle=\ttfamily\scriptsize,
  breaklines=true,
  breakatwhitespace=false,
  columns=fullflexible,
  frame=single,
  framesep=3pt,
  xleftmargin=3pt,
  xrightmargin=3pt,
  showstringspaces=false,
  commentstyle=\color{gray},
  keywordstyle=\bfseries,
  backgroundcolor=\color{black!3}
}
\newcommand{\framedshot}[2]{{\setlength{\fboxsep}{0pt}\setlength{\fboxrule}{0.4pt}%
  \fcolorbox{black!45}{white}{\includegraphics[width=\dimexpr#1-0.8pt\relax]{#2}}}}

\begin{document}

\title{Scriptorium: System Design and Implementation of a Multi-Human, Multi-AI Collaborative Digital Library Search Platform}
\subtitle{A Technical Report}

\author{Abbas Pirmoradi}
\orcid{0000-0001-6493-9285}
\affiliation{%
  \department{Department of Computer Science}
  \institution{University of Regina}
  \city{Regina}
  \country{Canada}}
\email{abbaspirmorady@uregina.ca}

\author{Milad Momeni}
\orcid{0000-0003-4267-4938}
\affiliation{%
  \department{Department of Computer Science}
  \institution{University of Regina}
  \city{Regina}
  \country{Canada}}
\email{miladmomeni@uregina.ca}

\author{Anna Chu}
\orcid{0009-0007-3724-5035}
\affiliation{%
  \department{Department of Computer Science}
  \institution{University of Regina}
  \city{Regina}
  \country{Canada}}
\email{annachu@uregina.ca}

\author{Orland Hoeber}
\authornote{Corresponding author.}
\orcid{0000-0003-2875-1132}
\affiliation{%
  \department{Department of Computer Science}
  \institution{University of Regina}
  \city{Regina}
  \country{Canada}}
\email{orland.hoeber@uregina.ca}

\renewcommand{\shortauthors}{Pirmoradi et al.}

\begin{abstract}
This technical report documents the design and implementation of Scriptorium, a digital library search interface designed to support \textit{multi-human, multi-AI collaboration}: full cross-user transparency of search activity combined with a Manager--Worker architecture of seven role-specialized generative AI agents. A companion short demonstration paper describes Scriptorium's interaction design; because a demonstration format is constrained to a few pages, it necessarily omits most implementation detail. This report documents Scriptorium's complete system architecture, data model, collaborative-interaction design rationale, and AI orchestration mechanism, including verbatim agent prompts, the deterministic guardrails that constrain them, the language model and retrieval configuration, and the steps needed to run the system, so that the design can be inspected, critiqued, reproduced, and reused by other researchers. We also discuss the system's current limitations and the user study we are planning to evaluate how the shared record of search activity and the scoped agents support collaborative search, including collaborators' awareness of one another's work and how they use the agents.
\end{abstract}



\maketitle

\section{Introduction}

Interactive information retrieval research has, over the last two decades, separately established that (1) complex academic search tasks benefit from collaborative support, awareness of collaborators' activity, mechanisms for dividing labour, and persistence of shared state~\cite{morris2007searchtogether}; and, more recently, that (2) such tasks benefit from generative-AI-augmented workspaces that scaffold an individual searcher's metacognitive planning, monitoring, and evaluation~\cite{choi2021orgbox, crescenzi2021metacognition, orin2025wesearch}. No existing system we are aware of combines both: a workspace in which \textit{several} people can see and build on one another's activity, and in which \textit{several}, differently scoped AI agents, rather than one general-purpose assistant, can be invoked to reason over that shared activity.

We built Scriptorium to explore this combination directly. A companion short demonstration paper describes Scriptorium's interaction design and its planned interactive demonstration~\cite{pirmoradi2027scriptorium}; that paper necessarily omits most implementation detail in order to fit a demonstration track's page limit. This report is the companion technical document referenced there. It is intended for readers who want to understand, reproduce, critique, or extend Scriptorium's architecture, rather than to argue for the research contribution itself, which is the demonstration paper's purpose.

The remainder of this report is organized as follows. Section~\ref{sec:related} situates Scriptorium's two design axes, collaborative transparency and multi-agent AI, within prior work, and traces each of the seven agents to its precedents. Section~\ref{sec:framing} describes the research framing and design goals behind the system. Section~\ref{sec:architecture} documents the software architecture, technology stack, and data model. Section~\ref{sec:collab} explains the design rationale behind Scriptorium's collaborative-interaction features. Section~\ref{sec:orchestrator} documents the multi-agent AI orchestrator in detail, including verbatim agent prompts, routing logic, and deterministic guardrails. Section~\ref{sec:repro} gives the language model, retrieval, and runtime configuration, the REST interface, and the steps needed to run the system. Section~\ref{sec:limitations} states the system's known limitations plainly. Section~\ref{sec:discussion} discusses the system's relationship to retrieval-augmented generation and outlines our planned evaluation.

\section{Related Work}
\label{sec:related}

\subsection{Collaborative Search}
SearchTogether~\cite{morris2007searchtogether} is the foundational system in this space, establishing that a collaborative Web search interface should provide \textit{awareness} (per-user query histories, page-visitation metadata, ratings and comments visible to teammates), \textit{division of labour} (integrated messaging, a recommendation mechanism, and automatic query-splitting across a group), and \textit{persistence} (an automatically generated shared summary of a session). Its evaluation is particularly instructive for Scriptorium's design: awareness features were rated the most valuable part of the system, while the automatic division-of-labour features (split search, multi-engine search) saw little use, in part because they implicitly required one group member to act as a leader pushing work onto others, a role that did not suit the peer collaborators in that study. Scriptorium's transparency-by-default workspace model can be read as pushing SearchTogether's awareness principle to its logical extreme: rather than requiring an explicit ``recommend'' or ``split search'' action to share information, \textit{all} activity within a workspace is visible to \textit{all} members continuously, without requiring anyone to take on a coordinating role.

\subsection{AI-Augmented Exploratory and Workspace-Based Search}
A separate line of work has embedded structured workspaces and, more recently, generative AI into exploratory search interfaces to support metacognition, the planning, monitoring, and evaluation of one's own search progress~\cite{choi2021orgbox, crescenzi2021metacognition}. Cross-session search, resuming a complex task after a gap of hours, days, or weeks, has been studied both as a behavioural phenomenon~\cite{li2020everyday} and as a design target, with visualized search-history timelines shown to support task reacquaintance~\cite{gomes2022crosssession, hoeber2026timelines}. WESearch~\cite{orin2025wesearch} extends this line with generative-AI-produced summaries at both a nested-workspace and whole-workspace level, explicitly framing this as \textit{workspace-augmented generation} (WAG) rather than retrieval-augmented generation (RAG), since the summarized content is exactly what a searcher has deliberately chosen to save, not what a retrieval ranker judged relevant to a query. We adopt this WAG framing in Scriptorium's own design (Section~\ref{sec:orchestrator}) and extend it to a setting where the workspace being summarized, critiqued, or bridged is the product of \textit{multiple} searchers' saved and discarded resources, not one person's. Accessibility-focused work such as SearchPath~\cite{seihoon2026searchpath} demonstrates a related but distinct use of a persistent timeline, as a working-memory aid for students with dyslexia, reinforcing a broader point in this literature that exploratory search interfaces should not assume uniform reading, memory, or attention capacity among their users~\cite{makri2021leveler}. Hoeber's design principles for exploratory search interfaces~\cite{hoeber2025principles} synthesize much of this body of work and directly informed several of Scriptorium's interaction choices, in particular the preference for lightweight, persistent, and interactive additions to a familiar search-engine-results-page paradigm over novel, complex visualizations. Prior work from our own group on dynamic tagging, highlighting, and filtering of saved search results~\cite{pirmoradi2024tagsr} and on keyword-linked exploratory browsing~\cite{hoeber2022klink} similarly motivated specific Scriptorium mechanisms (in particular, the timeline that can be filtered to show only saved or deleted results), though, again, all of this prior work is single-user. The rise of AI agents as active participants in the search process, rather than as a single conversational answer engine, is a recent framing~\cite{white2024agents} that Scriptorium operationalizes concretely through its seven-agent Manager--Worker architecture (Section~\ref{sec:orchestrator}), and which relates to, but is architecturally distinct from, retrieval-augmented generation approaches~\cite{gao2024rag} more broadly. Self-regulated learning during search, an individual's process of goal-setting, monitoring, and adapting strategy~\cite{urgo2024goalsetting}, becomes, in a multi-user workspace, a process that a group must additionally negotiate and reconcile, a phenomenon Scriptorium's transparency mechanisms are designed to surface rather than hide. Finally, motivating the underlying complexity Scriptorium is designed for, graduate student searchers report specific, persistent difficulty tracking what has already been searched and by whom, even when working entirely alone~\cite{hoeber2024gradstudent}, and idea-generation tools such as SearchIdea~\cite{chavula2023searchidea} demonstrate that AI-generated prompts, when properly scoped, can support creativity rather than merely convenience during academic search, a design lesson reflected in several of Scriptorium's individual agents.

\subsection{Precedents for the Seven Agents}
\label{sec:related-agents}
Each of Scriptorium's seven agents operationalizes a form of search support with an established research record. We summarize that record agent by agent, and then state what Scriptorium adds to it.

\textbf{HERMES (Navigator).} Query suggestion is among the longest-studied aids in interactive search. Kelly et al.~\cite{kelly2009suggestion} compared query and term suggestion features and found that searchers preferred suggested queries to suggested terms. In exploratory literature search, Medlar et al.~\cite{medlar2021suggestions} showed that alternative queries can also act as keyword summaries of the results page: with them, participants opened about half as many documents per query yet were exposed to over 40\% more results. Large language models can now write such queries; Wang et al.~\cite{wang2023boolean} found that ChatGPT generates high-precision Boolean queries for systematic review search, at a cost in recall. Closest to HERMES, Khakshoor and Hoeber~\cite{khakshoor2025genai} use generative AI to suggest broader and narrower queries grounded in a searcher's recent exploratory search activity. HERMES shares this grounding in search history but differs in whose history it reads: its suggestions are conditioned on the whole group's queries and on the titles of documents the group has saved \textit{and discarded}, so that a proposed query avoids ground any collaborator has already covered or rejected.

\textbf{JANUS (Bridge).} Two precedents frame JANUS. Literature-based discovery began with Swanson's observation that two literatures that had never cited one another, one linking dietary fish oil to reduced blood viscosity and one linking Raynaud's syndrome to high viscosity, jointly implied a hypothesis that neither contained~\cite{swanson1986fishoil}. In collaborative search, Pickens et al.~\cite{pickens2008mediation} introduced algorithmic mediation, in which a system combines the activity of two or more people searching together, simultaneously, so that one can survey for promising documents while the other judges them in depth. We return to how JANUS differs from both below.

\textbf{ARIA (Synthesizer).} Tombros and Sanderson~\cite{tombros1998summaries} showed that summaries tailored to the query let searchers judge relevance more accurately and more quickly than static summaries built from a document's leading sentences. Orin and Hoeber~\cite{orin2025wesearch} carried summarization into the workspace, generating summaries of what a searcher has saved rather than of what a ranker retrieved. ARIA narrows this further, summarizing only the documents a collaborator attaches to the current message.

\textbf{MINERVA (Methods).} Structured extraction from research abstracts has a long history in evidence-based medicine, where Nye et al.~\cite{nye2018pico} annotated 5{,}000 abstracts of randomized controlled trials with their populations, interventions, and outcomes to support automated extraction. With large language models, Khraisha et al.~\cite{khraisha2024llmreview} found that GPT-4 could rival humans on some systematic-review tasks under specific conditions, but that its data extraction was only moderate once chance agreement was accounted for. MINERVA is accordingly bounded to what an abstract states, and reports a missing element as not stated rather than inferring it.

\textbf{CLIO (Citations).} Walters and Wilder~\cite{walters2023fabrication} found that 55\% of the citations GPT-3.5 produced in short literature reviews, and 18\% of those produced by GPT-4, were fabricated, and that many citations to real works contained errors. CLIO therefore formats references from catalogue metadata rather than generating them, and marks a missing field instead of supplying one.

\textbf{SOCRATES (Critic).} Crescenzi et al.~\cite{crescenzi2021metacognition} designed a workspace tool to support the metacognitive planning and monitoring of exploratory search, and Danry et al.~\cite{danry2023askme} showed that AI explanations framed as questions improved people's logical discernment compared with explanations that simply asserted a conclusion. SOCRATES combines the two: it critiques the group's evidence base for gaps, and ends with a question rather than a verdict.

\textbf{MOMUS (Counterpoint).} Devil's advocacy is an established structure for group decisions: Schweiger et al.~\cite{schweiger1986devil} found that groups using devil's advocacy or dialectical inquiry produced higher-quality recommendations than groups seeking consensus. Chiang et al.~\cite{chiang2024devil} gave the role to a large language model and found that an LLM devil's advocate arguing against an AI recommendation can promote a group's appropriate reliance on AI without substantially increasing workload. The need is acute in search. White~\cite{white2013beliefs} showed that searchers favour positive information over negative, that search engines strongly favour one, usually positive, perspective irrespective of the truth, and that around half of the answers searchers settled on were incorrect; Sharma et al.~\cite{sharma2024echo} found that LLM-powered conversational search led participants to more biased querying than conventional search, especially when the model reinforced their views. MOMUS applies devil's advocacy to the corpus a group has assembled rather than to a single decision.

\textbf{What is new.} None of these capabilities is new on its own, and we do not claim otherwise. What is new is the setting in which all seven operate: one multi-author evidence base that includes what the group discarded, which HERMES, SOCRATES, and MOMUS read as negative relevance judgements. JANUS goes further: to our knowledge, it is a new kind of search support, and it differs from algorithmic mediation~\cite{pickens2008mediation} on four counts.
\begin{enumerate}
  \item \textit{Synchronous versus asynchronous.} Algorithmic mediation supports people searching at the same time. JANUS works over persistent activity accumulated across sessions, and is useful precisely when collaborators are not searching together.
  \item \textit{Retrieval versus interpretation.} Algorithmic mediation acts on retrieval, deciding which documents each partner is shown. JANUS acts on interpretation: it produces a natural-language account of what each collaborator's thread is about and where the threads connect, together with one bridging query to pursue that connection.
  \item \textit{Assigned roles versus peers.} Algorithmic mediation divides the work between complementary roles. JANUS assumes peers without roles, consistent with SearchTogether's finding that division-of-labour features which presume a leader went largely unused~\cite{morris2007searchtogether}.
  \item \textit{Inside the retrieval loop versus in the open.} Algorithmic mediation operates within the system's retrieval of results. JANUS is invoked on demand and posts its account into the shared chat, where any collaborator can read and contest it.
\end{enumerate}
Its input, the difference between collaborators' activity, has no single-user counterpart, which is why its preflight guard (Listing~\ref{lst:preflight}) refuses a request until at least two people have activity. Swanson's literature-based discovery~\cite{swanson1986fishoil} is the closer analogy in spirit, joining bodies of work that do not reference each other; JANUS applies that idea to the live, partial threads of a research group rather than to published literatures.

\section{Research Framing and Design Goals}
\label{sec:framing}

Scriptorium is a research prototype, not a production system. It deliberately makes search activity transparent across all users rather than siloed per account: any logged-in user can see and resume any workspace, view the full, colour-coded activity timeline of every collaborator, and read the shared chat thread, including the AI's responses within it. This is a design choice made in direct service of a research question, how full transparency of a group's search activity supports collaborative search, rather than a default that a production multi-tenant SaaS product would responsibly make. Four concrete design goals follow from this framing:

\begin{enumerate}
  \item \textbf{Ambient awareness over explicit sharing.} Following the SearchTogether finding discussed in Section~\ref{sec:related}, information about a collaborator's activity should be visible by default, not only when someone explicitly recommends or pushes it.
  \item \textbf{Auditable, not monolithic, AI assistance.} A group working together should be able to trust not only \textit{that} an AI agent produced an answer, but see \textit{what evidence, specifically}, it was allowed to see when producing it.
  \item \textbf{Genuine, not synthesized, activity records.} Every event surfaced in the collaborative timeline corresponds to a real, tracked system event (a query issued, a document saved, a document's removal flag flipped, a document opened); no event type is fabricated purely to make the timeline appear more complete, since the timeline itself may become a data source for studying the group's actual behaviour.
  \item \textbf{Lightweight extension of a familiar paradigm.} Consistent with Hoeber's design principles for exploratory search interfaces~\cite{hoeber2025principles}, collaborative and AI features are layered onto a conventional search-engine-results-page interaction model rather than replacing it with an unfamiliar paradigm.
\end{enumerate}

\section{System Architecture}
\label{sec:architecture}

\subsection{Technology Stack}
\begin{figure}[t]
  \centering
  \begin{tikzpicture}[
      node distance=6mm and 8mm,
      box/.style={draw, rounded corners, align=center, font=\footnotesize, minimum height=9mm, minimum width=30mm},
      >=Stealth
    ]
    \node[box] (fe) {React SPA (frontend)\\Redux Toolkit + Tailwind};
    \node[box, right=of fe] (be) {Express API (backend)\\Node.js, Mongoose};
    \node[box, right=of be] (db) {MongoDB\\(Atlas / local)};
    \node[box, below=of be, fill=black!5] (server-llm) {OpenAI API\\(\texttt{gpt-5})};
    \draw[<->] (fe) -- node[above,font=\tiny]{REST} (be);
    \draw[<->] (be) -- node[above,font=\tiny]{ODM} (db);
    \draw[<->] (be) -- (server-llm);
  \end{tikzpicture}
  \caption{Scriptorium architecture. The Manager--Worker orchestrator runs in the backend, which builds each agent's context from the database and calls the OpenAI API; the browser never contacts the model directly.}
  \Description{A block diagram showing a React single-page-application frontend connected via REST to an Express backend, which is connected via an object-document mapper to a MongoDB database. The backend also connects to the OpenAI API.}
  \label{fig:architecture}
\end{figure}
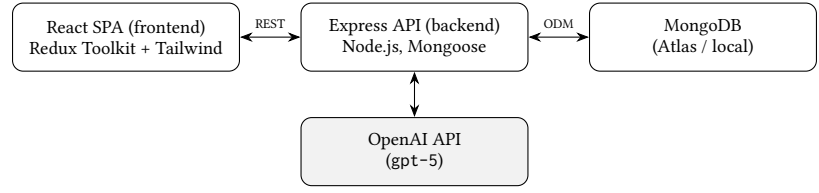

Scriptorium is built on the MERN stack (MongoDB, Express, React, Node.js)~\cite{Subramanian-2019}. The frontend is a React single-page application~\cite{hoque_2020} using Redux Toolkit for state management (one slice per data domain) and a combination of Tailwind CSS and hand-written CSS modules for an archival visual design system. The backend is an Express REST API running on Node.js~\cite{cantelonAndEtall-2014}, with Mongoose object-document mapping onto a MongoDB database~\cite{kristina_2013} (Atlas-hosted or local, depending on deployment). Figure~\ref{fig:architecture} summarizes the high-level architecture. The agent orchestrator (\texttt{backend/ai/AIOrchestrator.js}) runs on the server: the chat interface sends each agent request to \texttt{POST /api/ai/chat}, and the backend assembles the request's context from the database, applies the orchestrator's scoping and guards, and calls the OpenAI API (Section~\ref{sec:repro}). The model API key is therefore held only by the server and never reaches the browser. Document-search results shown on the search engine results page (SERP) are retrieved live from Ex Libris Primo~\cite{vaughan_2011}, an institutional library discovery and metadata API; Scriptorium is a collaborative \textit{layer} over an existing discovery system, not a replacement for one, which is what makes its mechanisms portable across any institution using a compatible discovery service.

\subsection{Data Model}
Table~\ref{tab:datamodel} summarizes the core Mongoose models. Two distinctions in this model are deliberate and worth stating explicitly. First, \texttt{Document} records distinguish \textit{saved} (\texttt{doc\_isRemoved: false}) from \textit{deleted} (\texttt{doc\_isRemoved: true}) resources via a soft-delete flag rather than a hard delete: a discarded resource remains visible in the collaborative timeline, dimmed and struck through, as a record of a direction the group considered and abandoned, which is itself a research-relevant signal distinct from a resource never having been found at all. Second, \texttt{ViewLog} records a genuine ``a user opened this document'' event, kept separate from \texttt{Document}, so that the activity timeline's four event facets (search, save, delete, view) each correspond to an actual tracked action rather than a synthesized one (see design goal 3, Section~\ref{sec:framing}).

\begin{table}[t]
  \small
  \caption{Core data model.}
  \label{tab:datamodel}
  \begin{tabular}{@{}p{0.16\linewidth}p{0.30\linewidth}p{0.46\linewidth}@{}}
    \toprule
    \textbf{Model} & \textbf{Purpose} & \textbf{Key fields} \\
    \midrule
    User & Account & username, password (bcrypt) \\
    Workspace & A named research project, owned but visible to all users & name, userId (owner), archive flag \\
    Query & One search-topic instance within a workspace & query text, userId, username, workspaceId, workspaceName \\
    Document & A saved (or since-removed) search result & documentId (catalogue record id), queryId, workspaceId, userId, username, queryKey, title, doc\_type, doc\_authors, doc\_date, doc\_abstract, doc\_url, doc\_isRemoved \\
    ChatMessage & One human or AI chat message & workspaceId, queryText, queryId, username, userIndex, text, agentKey, docRefs, suggestedQueries \\
    ViewLog & A genuine ``document opened'' event & workspaceId, queryId, documentId, username \\
    \bottomrule
  \end{tabular}
\end{table}

\subsection{Live-Update Mechanism}
The collaborative timeline and chat refresh via a three-second polling interval rather than WebSockets or server-sent events: the chat panel refetches the workspace's messages, and the results page refetches the workspace detail (every query with its saved and removed documents) and the view log, each every three seconds. A separate, slower poll (sixty seconds) refreshes the current topic's cross-user document list, which supplements the workspace detail when agent context is built (Section~\ref{sec:repro}). This was a deliberate simplicity choice: at the scale of a research prototype's target cohort (at most approximately nine concurrent testers in our planned study), a three-second poll is imperceptibly different from a push-based update, and avoids the operational complexity of running and maintaining a WebSocket server for a system that will not be deployed at production scale.

\section{Collaborative Interaction Design}
\label{sec:collab}

\begin{figure}[tbp]
  \centering
  \framedshot{\linewidth}{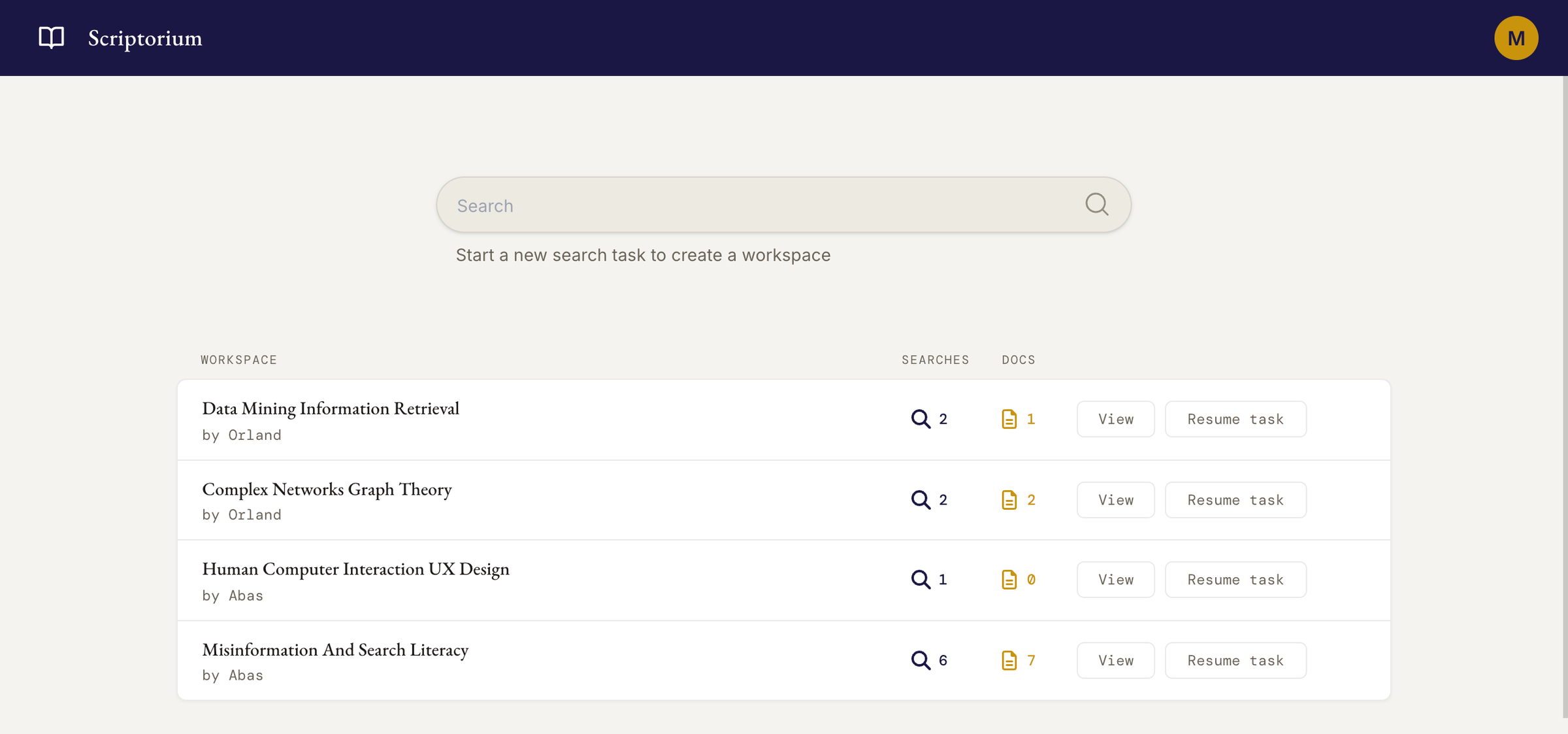}\\[1pt]
  {\small (a) Dashboard}\\[8pt]
  \framedshot{\linewidth}{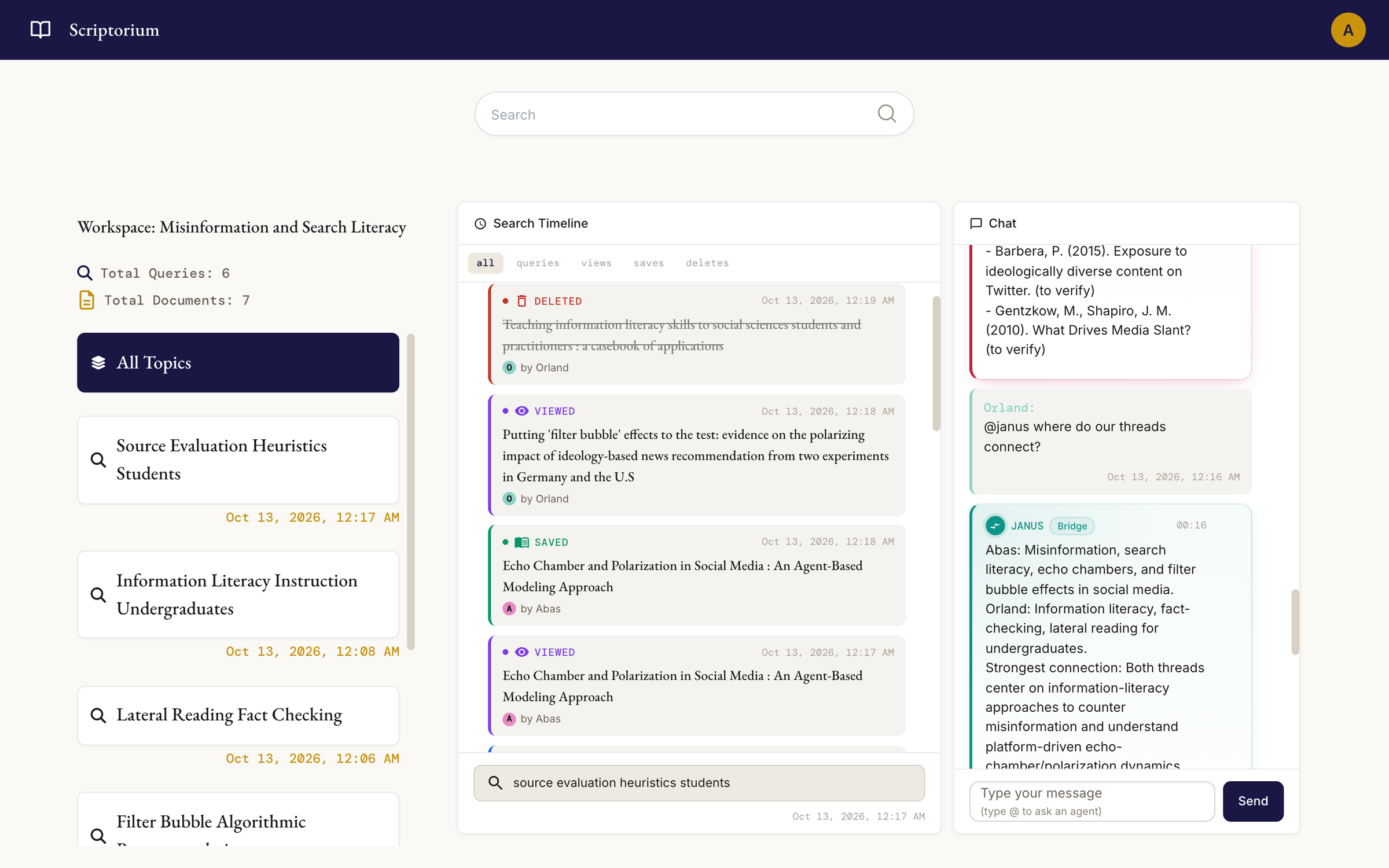}\\[1pt]
  {\small (b) Workspace details}
  \caption{The dashboard lists every workspace to every collaborator (a), and the workspace details page reviews a project's topics, timeline, and chat across sessions (b).}
  \Description{Two stacked screenshots. The top one shows the Scriptorium dashboard: a search box above a table of four workspaces, each with a title, a creator name, search and document counts, and View and Resume task buttons. The bottom one shows the workspace details page for the workspace Misinformation and Search Literacy: total query and document counts and a list of search topics on the left, then a Search Timeline with a struck-through deleted event, viewed events, and a saved event attributed to two collaborators, and beside it the workspace chat, whose latest message is a reply from the JANUS agent bridging two collaborators' threads.}
  \label{fig:screenshots-tr}
\end{figure}

\begin{figure}[tbp]
  \centering
  \framedshot{\linewidth}{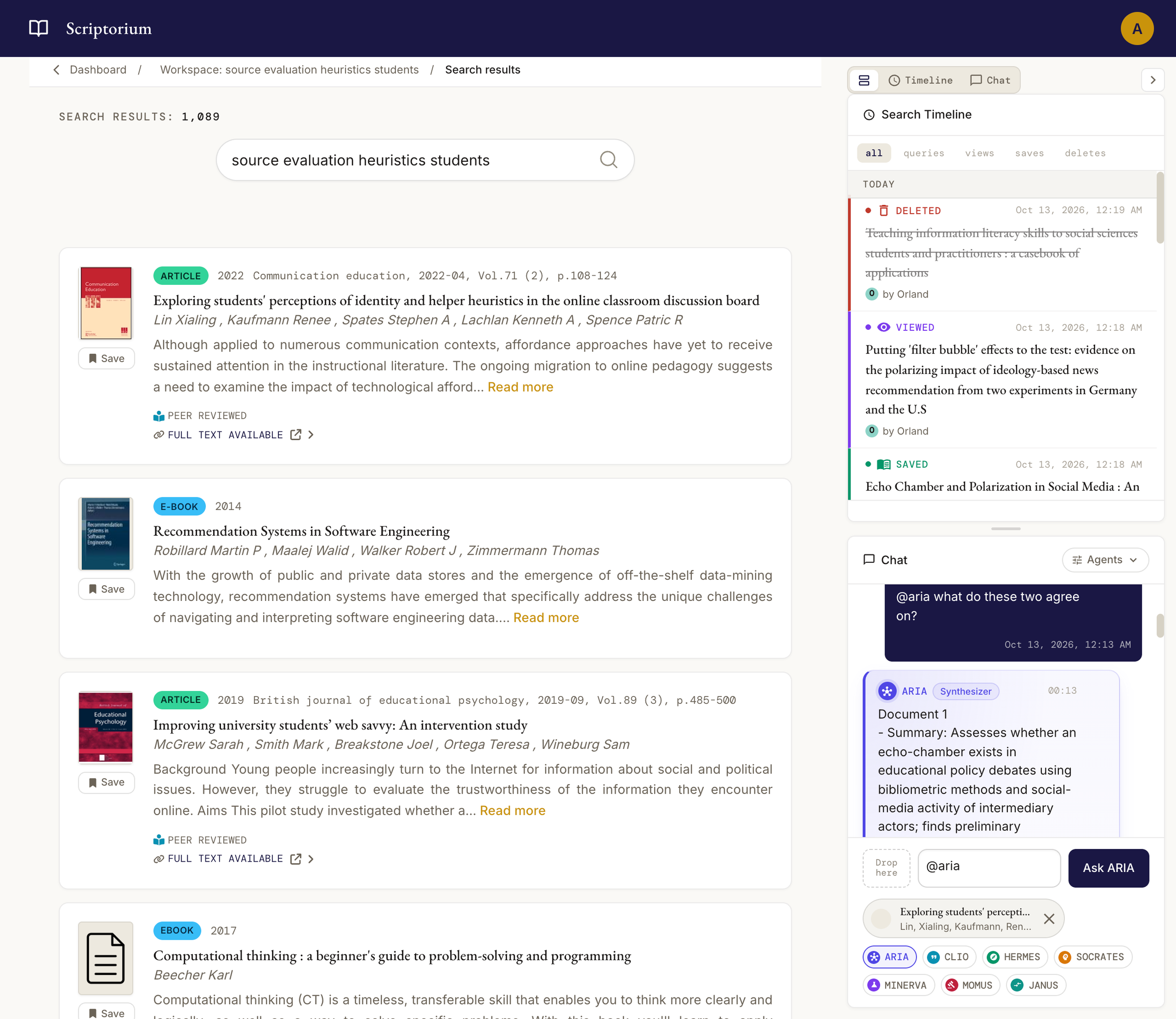}
  \caption{The SERP provides the library's search results alongside a dock that holds the group's activity timeline and shared chat.}
  \Description{A screenshot of the Scriptorium search results page for the query source evaluation heuristics students, showing three catalogue results with cover thumbnails, format badges, authors, and abstracts. A right-hand dock shows a Search Timeline with a struck-through deleted event, a viewed event, and a saved event by two collaborators, and below it a chat panel showing a request addressed to ARIA and ARIA's reply, above a message box holding a search result dragged in as an attachment, a message addressed to ARIA, and a row of agent chips.}
  \label{fig:serp-tr}
\end{figure}

Figures~\ref{fig:screenshots-tr} and~\ref{fig:serp-tr} show the three interface surfaces discussed in this section: the workspace dashboard, the workspace details page used for cross-session review, and the search results page with its collaborative dock.

\subsection{Transparent Workspaces}
Any logged-in user sees every workspace on the dashboard, a flat list sorted most-recent-first, not only workspaces they created or were invited to. Each card shows the creator's username, search and document counts, and two actions: \textit{View} (open the workspace's activity overview) and \textit{Resume task} (jump directly into the SERP for that workspace's most recent search). Edit, archive, and rename actions remain restricted to the workspace's creator, enforced in both the UI and, for the create/update path, server-side, so that ``everyone can view or resume'' is deliberately decoupled from ``everyone can modify.'' This is the system's central methodological choice: full cross-user visibility is what lets a collaborator find a teammate's search thread without being explicitly invited into it, and what makes such awareness observable at all.

\subsection{The Search Timeline}
A single chronological stream merges the four genuinely tracked event types described in Section~\ref{sec:architecture} (search, save, delete, view), filterable by type, grouped by day, and live-updating via the polling mechanism above. Two independent colour channels are used, deliberately kept visually disjoint so that a reader's eye never confuses \textit{what} happened with \textit{who} did it. The first channel encodes event type: the left border, icon, and type label of each timeline entry share one colour per event type (blue for search, emerald for save, red for delete, violet for view). These specific hues were not chosen arbitrarily: green and red for save and delete mirror the near-universal version-control diff convention (additions in green, removals in red), and violet for ``viewed'' mirrors the browser \texttt{:visited}-link convention dating to the earliest CSS specifications. The second channel encodes the acting user: a small coloured avatar-initial chip next to ``by \{username\}'' draws from a separate 15-colour palette (ColorBrewer ``Set3''), assigned first-come-first-served per session via a session-scoped username-to-colour map. This first-come-first-served assignment was chosen deliberately over a hash-based scheme, which was tried first and rejected, because hash collisions could silently assign two different users the same colour; the first-come-first-served map guarantees no two users share a colour for cohorts of up to fifteen concurrent participants.

\subsection{Persistent Chat}
A shared chat thread is scoped to a \textit{workspace} by default, persisting across however many distinct searches happen inside that workspace over its lifetime. An earlier version of the system scoped chat per search-topic instead; this caused chat history to visibly disappear the moment a new search was issued within the same workspace, a usability problem that was identified during internal testing and corrected by re-scoping chat to the workspace level. On the workspace-details page specifically (a page for reviewing past search topics one at a time via a sidebar), chat can optionally be re-scoped to whichever topic is currently selected in that sidebar, since that page's entire interaction model, unlike the active-search SERP, is built around browsing discrete past topics rather than a single continuous session.

\subsection{Adapting Interface Complexity to Task Context}
The same Timeline-and-Chat dock component renders two different ways depending on context. On the active-search SERP, it is collapsible to an icon strip, offers a combined mode with a resizable drag-divider between Timeline and Chat, and persists its configuration across sessions, on the premise that a researcher actively working a search benefits from maximum layout flexibility. On the workspace-details page, the same underlying component is instead always expanded and fills the space beside the topic sidebar. When that space is at least 690 pixels wide, Timeline and Chat sit side by side, keeping activity and conversation in view together as the SERP's combined mode does; narrower than that, the panel falls back to a two-way tab switch that opens on Timeline, and the hidden chat keeps polling so that its tab can show an unread count. The SERP's collapse and resize controls are omitted there, since that page already provides its own primary navigation (a sidebar of past topics) and the SERP's extra chrome would add complexity without benefit to a ``review what happened'' task rather than an ``actively searching'' one. Search-result pagination follows the same task-sensitive philosophy in a smaller way: rather than rendering every page number for result sets that can exceed 200 pages, or a sliding window that recenters on every click, results pages are shown in fixed blocks of up to ten page numbers, jumping to the next block of ten once a user pages past the current window, matching how large result sets are conventionally paginated.

\section{The Multi-Agent AI Orchestrator}
\label{sec:orchestrator}

\subsection{Design Rationale}
Rather than a single general-purpose assistant, Scriptorium implements seven named, specialized agents behind a Manager--Worker routing layer. The underlying design hypothesis is that giving each capability a distinct persona, a strictly scoped input, and a hard output-length limit produces responses that are more predictable, more citable, and more trustworthy to a group of collaborators than a single do-everything chatbot would, and additionally lets the interaction log itself become research data: which agent a group invoked, and at what point in their research process, is a meaningful behavioural signal that a single undifferentiated assistant would not expose.

The orchestrator runs on the server behind \texttt{POST /api/ai/chat} and reaches the model through a pluggable provider interface (\texttt{BaseLLMProvider}, a single \texttt{call(systemPrompt, userPrompt)} method), implemented for the OpenAI API. Earlier in development a second, hand-maintained copy ran in the browser against a free model proxy, and the two copies drifted: the browser copy at one point lacked two of the seven agents and passed every agent the same broad context rather than its scoped slice. Moving orchestration entirely to the server removed that copy, so that every listing in this section and in Appendix~\ref{sec:appendix-prompts} describes the only code that runs. Each request is first built into six context slices (Section~\ref{sec:repro}), from which the scoping function below selects.

\subsection{Agent Roster}
Table~\ref{tab:agents-full} lists all seven agents together with their strictly scoped inputs. The scoping is enforced in code, via a \texttt{buildScopedInput(agentKey, context)} function implemented as a \texttt{switch} statement that returns a different, minimal object per agent key (Listing~\ref{lst:scoping}), not by asking the underlying language model to self-restrict its attention to part of a larger context. For example, ARIA (the Synthesizer) receives only documents the user explicitly attached to the current chat message and never has access to the workspace's full saved-document corpus, while SOCRATES (the Critic) and MOMUS (the Counterpoint) receive the whole workspace's saved-and-discarded history because their purpose is specifically to critique or challenge the group's collective evidence base.

\begin{table}[t]
  \small
  \caption{The seven agents in Scriptorium's AI orchestrator.}
  \label{tab:agents-full}
  \begin{tabular}{@{}p{0.13\linewidth}p{0.13\linewidth}p{0.66\linewidth}@{}}
    \toprule
    \textbf{Agent} & \textbf{Role} & \textbf{Scoped input / behaviour} \\
    \midrule
    ARIA & Synthesizer & Documents attached to the current message only; summarizes each independently without cross-document comparison. \\
    CLIO & Citations & Attached documents' metadata, or saved documents if none attached; formats a reference list, marking missing metadata explicitly rather than inventing it. \\
    HERMES & Navigator & Query history plus saved/deleted titles; returns five clickable, $\leq$7-word query suggestions as structured JSON. \\
    SOCRATES & Critic & Whole workspace: all queries, saved and deleted documents, recent chat; identifies evidence gaps and asks one sharpening question. \\
    MINERVA & Methods & Attached (or saved) documents' abstracts; extracts study design, data source, sample, and analysis method where stated. \\
    MOMUS & Counterpoint & Saved/deleted corpus, query history, recent chat; states the corpus's dominant lean and surfaces contradictions and counter-evidence. \\
    JANUS & Bridge & Per-collaborator activity across the whole workspace; connects two or more collaborators' separate research threads and proposes one clickable, $\leq$7-word bridging query as structured JSON. \\
    \bottomrule
  \end{tabular}
\end{table}

\begin{lstlisting}[caption={Per-agent input scoping, condensed from \texttt{AIOrchestrator.js}. Each branch constructs a fresh object; fields not named here are never sent to that agent's model call.},label={lst:scoping}]
buildScopedInput(agentKey, context) {
    const titlesOnly = (docs) =>
        docs.map((d) => ({ title: d.title }));
    const metaOnly = (docs) => docs.map((d) => ({
        title: d.title, authors: d.authors || "",
        date: d.date || "" }));
    const withAbstract = (docs) => docs.map((d) => ({
        title: d.title, abstract: d.abstract || "" }));
    const conversation = (context.conversation || []).slice(-8);

    switch (agentKey) {
        case "aria":
            return { attachedDocs: withAbstract(attachedDocs),
                     conversation };
        case "clio":
            return { docs: metaOnly(attachedDocs.length > 0
                     ? attachedDocs : savedDocs), conversation };
        case "hermes":
            return { queryHistory,
                     savedDocs: titlesOnly(savedDocs),
                     deletedDocs: titlesOnly(deletedDocs) };
        case "socrates":
            return { queryHistory,
                     savedDocs: withAbstract(savedDocs),
                     deletedDocs: withAbstract(deletedDocs),
                     conversation };
        case "minerva":
            return { docs: withAbstract(attachedDocs.length > 0
                     ? attachedDocs : savedDocs), conversation };
        case "momus":
            return { savedDocs: withAbstract(savedDocs),
                     deletedDocs: withAbstract(deletedDocs),
                     queryHistory, conversation };
        case "janus":
            return { activityByUser, conversation };
        default:
            return { conversation };
    }
}
\end{lstlisting}

Three properties of Listing~\ref{lst:scoping} are worth drawing out. First, the three projection helpers in that listing mean that scoping operates at field granularity, not only at document granularity: HERMES is given the \textit{titles} of saved and deleted documents so that it can avoid re-suggesting ground the group has already covered, but is never given their abstracts, because query reformulation does not require them. Second, the chat transcript is truncated to the last eight turns for every agent that receives it, and HERMES receives none at all. Third, CLIO and MINERVA implement a documented fallback (attached documents if any, otherwise the workspace's saved corpus), which is the only place where an agent's input set is state-dependent rather than fixed; both agents' preflight guards refuse the request when both sets are empty.

\subsection{Deterministic Preflight Guards}
Before any language-model call is made, each agent request passes through a deterministic preflight check that can refuse the request outright based on the state of the workspace, rather than relying on the model itself to recognize and report that it lacks sufficient input. Listing~\ref{lst:preflight} reproduces this guard verbatim from the running system.

\begin{lstlisting}[caption={Deterministic preflight guards (verbatim, \texttt{AIOrchestrator.js}).},label={lst:preflight}]
preflight(agentKey, context) {
    const attached = context.attachedDocs || [];
    const saved = context.savedDocs || [];
    const activity = context.activityByUser || {};

    if (agentKey === "aria" && attached.length === 0) {
        return "ARIA summarizes exactly what you attach. Drag one or
        more documents into the chat, then ask again.";
    }
    if ((agentKey === "clio" || agentKey === "minerva")
        && attached.length === 0 && saved.length === 0) {
        return `${this.agents[agentKey].name} needs documents.
        Attach some, or save documents to this topic first.`;
    }
    if ((agentKey === "socrates" || agentKey === "momus")
        && saved.length === 0
        && (context.deletedDocs || []).length === 0) {
        return `${this.agents[agentKey].name} works on this
        workspace's saved documents - there are none yet.
        Save a few papers first.`;
    }
    if (agentKey === "janus"
        && Object.keys(activity).length < 2) {
        return "JANUS bridges collaborators - it needs
        search/save activity from at least two people
        in this workspace.";
    }
    return null;
}
\end{lstlisting}

\subsection{Routing}
Routing is hybrid. An explicit \texttt{@agentname} mention in a chat message bypasses the manager entirely and delegates directly. Otherwise, a manager persona classifies the free-text request, calling an LLM with a short, rule-based routing prompt (Listing~\ref{lst:routing}) whose user content is the message alone (\texttt{\{"userMessage": ...\}}), so that no workspace content is exposed at the routing step; the call returns structured JSON indicating either a delegation target and a confidence score, or a request to clarify. Below a confidence threshold (0.7 in the current implementation), the manager asks a clarifying question listing all active agents rather than guessing, so that a low-confidence routing decision never masquerades as a confident one. If the LLM routing call itself fails, the system falls back to a keyword-based heuristic classifier rather than crashing or silently guessing.

\begin{lstlisting}[caption={Manager routing prompt (verbatim excerpt).},label={lst:routing}]
You are the Research Manager. Route the user's request to
EXACTLY ONE worker, or clarify.
Routing rules:
- summarize/compare attached docs -> aria
- citations/references/bibliography -> clio
- better/new search queries, keywords -> hermes
- gaps, what's missing, critique evidence -> socrates
- methodology, study design, how studies were done -> minerva
- contradictions, counter-evidence, challenge our
  assumption, devil's advocate -> momus
- connect/bridge collaborators' work,
  who is researching what -> janus
- If confidence < 0.7, return clarify.
Return ONLY JSON:
{"action":"delegate","agent":"<key>","conf":0.9}
or {"action":"clarify","conf":0.4}
\end{lstlisting}

\subsection{Prompt Design and Hard Output Limits}
Every agent's instruction string states its exact expected input object (e.g., \texttt{input.attachedDocs}, \texttt{input.queryHistory}) and a hard output-length limit, either a word count or a per-item line limit, stated directly in the prompt. Five of the seven agents rely on the language model to respect this stated limit. HERMES and JANUS are the exceptions: because their structured JSON output feeds a UI control (individually clickable query-suggestion chips) that would break on malformed input, their queries are additionally clamped in code (a \texttt{clampQueryWords} function enforcing a maximum of seven words per query, and at most five queries from HERMES and one from JANUS) rather than trusted entirely to the model. JANUS's reply text is likewise assembled in code from separate JSON fields (an optional verdict, one line per collaborator's thread, the bridge, and a work to verify), so its layout does not depend on the model following formatting instructions. This asymmetry, prompt-only limits for free-text agents, code-enforced limits for the two agents with a machine-consumed output, reflects a general principle in Scriptorium's AI integration: constraints that affect only a human reader's experience are prompted; constraints that would break a downstream UI component if violated are additionally enforced in code.

\section{Reproducibility}
\label{sec:repro}

This section collects the configuration a reader needs to run Scriptorium and to interpret its agents' behaviour.

\subsection{Language Model}
Every model call, routing and worker alike, is made by the backend through the OpenAI Chat Completions API, which the provider \texttt{backend/ai/OpenAILLMProvider.js} calls directly with Node.js's built-in \texttt{fetch}, with no client library. The model is \texttt{gpt-5}~\cite{openai2025gpt5}, configurable through the \texttt{OPENAI\_MODEL} environment variable, and the provider sets no sampling or reasoning parameters, so the API's defaults apply. Each call sends two messages:
\begin{lstlisting}
system: <agent instruction, Appendix A>
user:   <JSON: {"userRequest": ..., "input": <scoped slice>}>
\end{lstlisting}
\noindent For routing calls the user message is \texttt{\{"userMessage": ...\}} instead (Section~\ref{sec:orchestrator}). A request therefore costs one model call when addressed to an agent by name, and two (routing, then the worker) when delegated with \texttt{@ai}; refusals by a preflight guard cost none. When no API key is configured, \texttt{POST /api/ai/chat} returns an HTTP 503 error stating that the AI is not configured, rather than falling back to a stub. Substituting another model or vendor requires implementing only \texttt{call(systemPrompt, userPrompt)}.

\subsection{Context Construction}
Before scoping, each agent request is assembled into six slices by \texttt{buildWorkspaceContext} in \texttt{backend/controllers/aiController.js}, which reads the workspace's queries and documents directly from the database rather than trusting the client's copy:
\begin{itemize}
  \item \texttt{attachedDocs}: documents the user dragged into the current message, flattened from the catalogue record to title, abstract, authors, date, and type;
  \item \texttt{queryHistory}: every query issued in the workspace by any collaborator, oldest first, with its author;
  \item \texttt{savedDocs} and \texttt{deletedDocs}: every document saved in the workspace, split on the soft-delete flag and de-duplicated by catalogue record identifier, so that a paper saved by two collaborators appears once;
  \item \texttt{activityByUser}: for each collaborator, the queries they issued and the titles of the documents they currently keep saved;
  \item \texttt{conversation}: the last twelve chat turns, each with speaker, an AI flag, and text, of which the scoping function retains the last eight.
\end{itemize}
The scope of these slices is the whole workspace rather than the current search topic, so that an agent invoked during one search can reason over the group's work on every other topic in the same project.

\subsection{Retrieval}
Search results are requested from the Ex Libris Primo Search REST API (\texttt{GET /primo/v1/search}) with the query string passed as \texttt{q=any,contains,<query>}, the institution's \texttt{inst} and \texttt{vid} parameters, \texttt{tab=Everything}, \texttt{scope=MyInst\_and\_CI} (local holdings plus the Central Discovery Index), and \texttt{offset}/\texttt{limit} for pagination. Records are consumed in Primo's PNX format; the fields Scriptorium uses are \texttt{pnx.display.title}, \texttt{pnx.display.type}, \texttt{pnx.display.creationdate}, \texttt{pnx.addata.au}/\texttt{addau} (authors), and \texttt{pnx.addata.abstract}. Cover images are resolved by DOI through the Third Iron (BrowZine) public API and by ISBN through Syndetics. Retrieval is therefore not reproducible bit-for-bit across institutions: a deployment at another institution needs its own Primo API key and view identifiers, and will return that institution's licensed collection.

\subsection{REST Interface}
Table~\ref{tab:api} lists the endpoints the interface uses. All are served by the Express backend under \texttt{/api}.

\begin{table}[t]
  \small
  \caption{REST endpoints used by the interface.}
  \label{tab:api}
  \begin{tabular}{@{}p{0.44\linewidth}p{0.50\linewidth}@{}}
    \toprule
    \textbf{Endpoint} & \textbf{Purpose} \\
    \midrule
    \texttt{POST /auth/register}, \texttt{/auth/login} & Create an account; sign in \\
    \texttt{GET /workspaces} & List every workspace (dashboard) \\
    \texttt{GET /workspaces/:workspaceId} & Workspace detail: queries with saved and removed documents \\
    \texttt{POST /workspaces}, \texttt{PUT|DELETE /workspaces/:id} & Create; rename or archive, delete (creator only) \\
    \texttt{POST /workspaces/query} & Record a search query \\
    \texttt{POST /workspaces/document} & Save a result to a query \\
    \texttt{PATCH /workspaces/document/toggle/:documentId} & Flip the soft-delete flag \\
    \texttt{GET /workspaces/document/query/:queryId} & Cross-user documents for a topic \\
    \texttt{POST /workspaces/document/view} & Log a document-opened event \\
    \texttt{GET /workspaces/document/views/:workspaceId} & View events for the timeline \\
    \texttt{GET|POST /chat}, \texttt{DELETE /chat/:messageId} & Read, post, remove chat messages \\
    \texttt{GET /ai/agents}, \texttt{POST /ai/chat} & Agent registry; server-side orchestration \\
    \bottomrule
  \end{tabular}
\end{table}

\subsection{Running the System}
The repository contains the backend (\texttt{backend/}, Node.js with Express~4 and Mongoose~8) and the frontend (\texttt{frontend/}, React~18 created with Create React App~5, Redux Toolkit~1.9, Tailwind CSS~3, and Material UI~7). Mongoose~8 requires Node.js~16 or later. The backend reads \texttt{MONGO\_URI}, \texttt{OPENAI\_API\_KEY}, and, optionally, \texttt{OPENAI\_MODEL} (default \texttt{gpt-5}) and \texttt{PORT} (default 5006) from a \texttt{.env} file in the repository root (template: \texttt{backend/.env.example}); the frontend reads the institution's Primo API key from \texttt{REACT\_APP\_PRIMO\_API\_KEY} in \texttt{frontend/.env.local}, and the backend address and the remaining Primo parameters from \texttt{frontend/src/config/env.js}. The frontend template is \texttt{frontend/.env.example}. After \texttt{npm install} in the repository root, in \texttt{backend/}, and in \texttt{frontend/}, \texttt{npm run dev} in the root starts both processes; with \texttt{NODE\_ENV=production} the backend also serves the built frontend (\texttt{npm run build} in \texttt{frontend/}). Running the agents requires an OpenAI API key with access to the configured model; usage is billed to that key's account.

\subsection{Code Availability}
The source code is available at \url{https://github.com/abbaspirmoradi/Scriptorium-DL-Interface} under the MIT licence.

\section{Known Limitations}
\label{sec:limitations}

We state these plainly, consistent with the goal of this report as a companion technical document rather than a promotional one.

\textbf{No production-grade authentication.} There is no signed session token (no JWT); the client stores a plain \texttt{\{\_id, username\}} object in \texttt{localStorage}, and every backend endpoint trusts whatever identifier the client sends. This is acceptable for a controlled study among cooperative, consenting participants, to whom the property is disclosed, but would require real authentication before any adversarial or production deployment.

\textbf{Dependence on a hosted model.} The agents depend on a commercial model served by OpenAI: the snapshot behind the \texttt{gpt-5} alias, its defaults, its pricing, and its availability are outside our control and may change over time. A reader reproducing agent behaviour at a later date should pin a dated model snapshot through \texttt{OPENAI\_MODEL}, and should expect outputs to differ for this reason as well as the next. Workspace content sent to agents, including titles, abstracts, queries, and recent chat, is transmitted to OpenAI, which participants in a study must be told.

\textbf{Polling, not push.} Chat and the activity timeline refresh via a three-second \texttt{setInterval} poll rather than WebSockets or server-sent events, simple and reliable at the intended small scale, but not the mechanism a production-scale version of this system would use.

\textbf{Model non-determinism.} Because agent responses are generated by a language model, two invocations of the same agent on the same workspace state may produce different, though similarly scoped, output. Hard output-length limits and precise, narrowly scoped prompts (Section~\ref{sec:orchestrator}) reduce, but do not eliminate, the practical impact of this non-determinism.

\section{Discussion and Planned Evaluation}
\label{sec:discussion}

Scriptorium's AI orchestrator is, in the terms proposed by Orin and Hoeber~\cite{orin2025wesearch}, a workspace-augmented generation (WAG) system rather than a retrieval-augmented generation (RAG) one: every agent's output is grounded only in resources that one or more human collaborators have explicitly saved (or, for SOCRATES and MOMUS, explicitly discarded), not in documents a retrieval ranker judged relevant to a query. We extend their single-user framing of WAG to a multi-user one: the workspace an agent reasons over in Scriptorium is the joint product of a group's saved and discarded resources, and at least one agent, JANUS, is defined specifically in terms of the difference between two or more collaborators' individual activity within that shared workspace, a construct that has no single-user analogue.

We are in the process of designing a controlled user study, structured similarly in spirit to the SearchTogether evaluation~\cite{morris2007searchtogether}, in which pairs or small groups of participants complete a realistic, multi-session academic search task using Scriptorium. Beyond standard search-interface metrics, we intend to specifically examine (1) transactive memory, whether and how quickly a collaborator becomes aware of a teammate's relevant prior activity without being explicitly told about it, comparing workspaces where the collaborative timeline is available against a condition where it is not; (2) which of the seven AI agents groups invoke, and at what point in the exploratory-to-focused search transition~\cite{white2009exploratory} they invoke it; and (3) whether the auditable, per-agent input scoping described in Section~\ref{sec:orchestrator} measurably increases participants' stated trust in AI-generated content relative to a single undifferentiated assistant baseline, an axis of comparison the companion demonstration's interactive session is also designed to probe informally with conference attendees.

\section{Conclusion}
We have documented the complete system architecture, data model, collaborative-interaction design rationale, and multi-agent AI orchestration mechanism behind Scriptorium, a digital library search interface built to support multi-human, multi-AI collaboration. This report is intended to make the system's design inspectable and reusable independent of the page constraints of its companion demonstration paper~\cite{pirmoradi2027scriptorium}, and we welcome scrutiny of, and extensions to, the mechanisms described here, particularly the deterministic preflight guards and per-agent input scoping approach, which we believe generalize beyond the specific context of digital library search to other collaborative sensemaking tools that incorporate generative AI.

\begin{acks}
This work was supported by a University of Regina President's Research Seed Grant for Health, Science, and Engineering.

This research was conducted at the University of Regina, which is situated on Treaty 4 and Treaty 6 territories, the homelands of the n\^ehiyawak, Anih\v{s}inap\=ek, Dakota, Lakota, Nakoda peoples, and Michif/M\'etis nation.
\end{acks}


\appendix

\section{Verbatim Agent Instruction Strings}
\label{sec:appendix-prompts}

This appendix reproduces the instruction string of all seven workers, supplementing Section~\ref{sec:orchestrator}. Each is the complete system prompt passed as the first argument to \texttt{llm.call(agent.instruction, JSON.stringify(workerInput))}, where \texttt{workerInput} carries the user's message under \texttt{userRequest} and the scoped object of Listing~\ref{lst:scoping} under \texttt{input}. The prompts are reproduced unmodified from the running system's source (\texttt{backend/ai/AIOrchestrator.js}), except that lines are re-wrapped to the column width, and em dashes, arrows, and the \texttt{$\leq$} symbol are rendered as \texttt{-}, \texttt{->}, and \texttt{<=} for typesetting. Each is stored in source as a concatenation of \texttt{\textbackslash n}-terminated string literals; the line breaks shown are therefore the ones the model actually receives, apart from re-wrapping.

Three conventions recur across all seven and are the substance of the design described in Section~\ref{sec:orchestrator}: a named persona and role in the opening line; an explicit \texttt{INPUT:} declaration naming the exact fields the agent will receive, closed by the sentence \textit{Nothing else exists}; and a closing \texttt{HARD LIMIT:} stating a word count or a per-item line budget. Six agents, all but HERMES, receive \texttt{input.conversation}; ARIA, SOCRATES, MOMUS, and JANUS additionally carry a rule constraining what the transcript may be used for, so that recent chat serves as referent resolution rather than as an unbounded second corpus. CLIO's and MINERVA's prompts carry no such rule and do not mention the transcript, so for those two agents its use is left to the model.

\begin{lstlisting}[caption={ARIA (Synthesizer). Its \texttt{INPUT} is the narrowest of the seven, and its preflight guard refuses outright when \texttt{attachedDocs} is empty.}]
You are ARIA, the Synthesizer.
INPUT: input.attachedDocs - the documents the user attached
to this message. Nothing else exists.
TASK: Summarize EXACTLY these documents - one short summary
per document plus a one-line takeaway. Do NOT compare or
contrast them against each other.
RULES:
- Never reference documents outside input.attachedDocs.
- Use input.conversation only to resolve follow-up phrases
  like 'the second one'.
- No preamble. Start with the substance.
HARD LIMIT: 120 words.
\end{lstlisting}

\begin{lstlisting}[caption={CLIO (Citations). The prohibition on inventing a DOI, page range, or publisher is the agent-level counterpart to the field-level restriction that CLIO receives metadata only, never abstracts.}]
You are CLIO, the Citations specialist.
INPUT: input.docs - document metadata (title, authors,
date). Nothing else exists.
TASK: Produce a formatted reference list (APA 7th unless
the user names another style).
RULES:
- Mark missing metadata as [n.d.] / [publisher unknown].
  NEVER invent a DOI, page range, or publisher.
- One citation per line. No commentary between citations.
HARD LIMIT: the reference list plus at most one closing
sentence.
\end{lstlisting}

\begin{lstlisting}[caption={HERMES (Navigator). One of the two agents, with JANUS, whose output is machine-consumed; both the five-query count and the seven-word cap are re-enforced in code after the call (Section~\ref{sec:orchestrator}).}]
You are HERMES, the Navigator.
INPUT: input.queryHistory (past searches), input.savedDocs
and input.deletedDocs (titles only). Nothing else exists.
TASK: Propose EXACTLY 5 new search queries that extend or
sharpen this workspace's search direction. Avoid repeating
past queries. Each query MUST be 7 words or fewer.
OUTPUT: Return ONLY this JSON, no other text:
{"note":"<one sentence, max 15 words>",
 "queries":["q1","q2","q3","q4","q5"]}
\end{lstlisting}

\begin{lstlisting}[caption={SOCRATES (Critic). Deleted documents are supplied with abstracts rather than titles alone, so that a direction the group considered and dropped remains available for critique.}]
You are SOCRATES, the Critic.
INPUT: input.queryHistory, input.savedDocs, input.deletedDocs
(titles + abstracts), input.conversation (recent chat turns).
Nothing else exists.
TASK: If the user is asking you to judge a specific claim
(their own or a named collaborator's, e.g. 'is X right?'),
find that claim in input.conversation, state directly whether
the evidence base supports it, and cite what's missing that
would change your answer. Otherwise, list the most important
gaps in this evidence base - missing themes, methods,
populations, or perspectives. Treat deleted documents as
directions the group considered and dropped; flag if a
dropped direction left a gap.
RULES:
- Use input.conversation only to resolve who said what and
  what they claimed - never invent a claim that isn't there.
- Numbered list, max 4 gaps, one line each.
- End with ONE pointed question that would sharpen the
  research direction.
HARD LIMIT: 100 words.
\end{lstlisting}

\begin{lstlisting}[caption={MINERVA (Methods). The qualifier \textit{only where stated or clearly inferable from the abstract} bounds the agent to the evidence actually supplied, since full texts are not retrieved.}]
You are MINERVA, the Methods specialist.
INPUT: input.docs - titles + abstracts. Nothing else exists.
TASK: For each document, extract: design, data source,
sample, analysis method - only where stated or clearly
inferable from the abstract.
RULES:
- One compact line per document: 'Title - design; data;
  sample; analysis'. Write 'not stated' for missing parts.
  NEVER invent details.
- If asked to compare, add one closing line grouping by
  methodology family.
HARD LIMIT: one line per document + one closing line.
\end{lstlisting}

\begin{lstlisting}[caption={MOMUS (Counterpoint). Step 4 is the one place in the roster where an agent is licensed to draw on knowledge outside the workspace, and it is paired with a mandatory 'to verify' label.}]
You are MOMUS, the Counterpoint. Your job is to counter
confirmation bias.
INPUT: input.savedDocs and input.deletedDocs (the group's
consensus corpus, titles + abstracts), input.queryHistory,
input.conversation (recent chat turns). Nothing else exists.
TASK:
1. If the user is asking you to judge a specific person's
   claim (e.g. 'is X right?'), find that claim in
   input.conversation and open with a direct verdict -
   supported, contradicted, or unsupported by this corpus -
   before anything else.
2. State in ONE sentence the dominant claim/assumption this
   corpus leans toward.
3. List any contradictions BETWEEN the provided documents
   (cite their titles).
4. Name up to 3 well-known opposing works, authors, or
   counter-arguments from scholarly knowledge that challenge
   the dominant claim - clearly label these 'to verify'
   since you cannot fetch them.
RULES:
- Use input.conversation only to resolve who said what -
  never invent a claim that isn't there.
- Be adversarial but factual. Never soften into agreement.
- If the corpus is too small or balanced to show a lean,
  say so in one sentence and stop.
HARD LIMIT: 120 words.
\end{lstlisting}

\begin{lstlisting}[caption={JANUS (Bridge). The only agent whose input is organized by collaborator rather than by document, and the only one gated on a property of the group rather than of the corpus.}]
You are JANUS, the Bridge. You connect collaborators'
separate research threads (transactive memory: who is
looking at what).
INPUT: input.activityByUser - each collaborator's queries
and saved-document titles, grouped by username - and
input.conversation (recent chat turns). Nothing else exists.
FIELDS:
- verdict: only if the user asks you to judge a specific
  person's claim (e.g. 'is X right?'), find that claim in
  input.conversation, check it against that person's
  activityByUser entry, and give the verdict in one
  sentence; otherwise an empty string.
- threads: one entry per collaborator, naming in one
  sentence what their thread is about.
- bridge: one sentence naming a specific concept,
  mechanism, or open question that links two (or more)
  threads, not the topic they already share.
- toVerify: ONE known work that addresses both threads,
  if scholarly knowledge suggests one; otherwise an empty
  string.
- queries: ONE bridging search query of 2-7 catalogue
  keywords that pursues the bridge, with no filler words
  such as 'bridge', 'research', 'studies', or 'and'.
RULES:
- Use input.conversation only to resolve who said what -
  never invent a claim that isn't there.
OUTPUT: Return ONLY this JSON, no other text:
{"verdict":"","threads":[{"user":"<username>",
 "about":"<one sentence>"}],"bridge":"<one sentence>",
 "toVerify":"<work or empty>","queries":["<query>"]}
\end{lstlisting}

\section{Agent Resolution and Manager Fallback}
\label{sec:appendix-routing}

An explicit mention is resolved before any model call, and each agent accepts a set of role aliases in addition to its name, so that a collaborator who remembers a function but not a persona still reaches the right worker: \texttt{@aria} also answers to \texttt{synthesizer}, \texttt{summarizer}, \texttt{summary}, \texttt{sum}; \texttt{@clio} to \texttt{citations}, \texttt{cite}, \texttt{citation}; \texttt{@hermes} to \texttt{navigator}, \texttt{reformulator}, \texttt{reformulate}, \texttt{queryhelper}; \texttt{@socrates} to \texttt{critic}, \texttt{gapdetector}, \texttt{gap}, \texttt{gaps}; \texttt{@minerva} to \texttt{methods}, \texttt{methodology}; \texttt{@momus} to \texttt{devil}, \texttt{devilsadvocate}, \texttt{contradict}, \texttt{counter}; and \texttt{@janus} to \texttt{bridge}, \texttt{connect}, \texttt{synthesisbridge}. The mentions \texttt{@ai} and \texttt{@manager} address the manager and therefore trigger routing rather than direct delegation.

Should the manager's routing call fail, a keyword classifier substitutes for it, tested in a fixed order so that the more specific collaborative intents are matched before the general ones. Terms including \textit{contradict}, \textit{refute}, \textit{devil}, and \textit{bias} route to MOMUS; \textit{bridge}, \textit{overlap}, \textit{teammate}, \textit{collaborator}, and \textit{who is looking} route to JANUS; \textit{gap} and \textit{missing} to SOCRATES; citation and method terms to CLIO and MINERVA respectively; \textit{summar}, \textit{overview}, and \textit{compare} to ARIA; and query-reformulation terms such as \textit{keyword}, \textit{boolean}, \textit{broaden}, and \textit{narrow} to HERMES. Every branch is additionally conditioned on the agent being enabled in the workspace's agent roster, and an unmatched request returns the clarification response listing the active agents rather than a guess. The heuristic is a degraded mode, not a design goal; we report it because a reader reproducing the system will otherwise observe routing behaviour that the manager prompt alone does not explain.

\end{document}